\documentclass[aps,prl,twocolumn]{revtex4}
\usepackage{graphicx}
\usepackage{amssymb}
\usepackage{amsfonts}
\usepackage{amsmath}
\usepackage{dcolumn}
\usepackage{natbib}

\begin{document}
\title{First Principles Study of Bismuth Films at Transition Metal Grain Boundaries}
\author{Qin Gao and Michael Widom}
\affiliation{Department of Physics, Carnegie Mellon University, Pittsburgh, Pennsylvania 15213, USA}

\begin{abstract}

Recent experiments suggest that Bi impurities segregate to form bilayer films on Ni and Cu grain boundaries but do not segregate in Fe. To explain these phenomena, we study the total energies of Bi films on transition metal (TM) $\Sigma$3(111) and $\Sigma$5(012) GBs using density functional theory. Our results agree with the observed stabilities. We propose a model to predict Bi bilayer stability at Ni GBs which suggests that Bi bilayer is not stable on (111) twist CSL GBs but is stable in most (100) twist CSL GBs. We investigate the interaction and bonding character between Bi and TMs to explain the differences among TMs based on localization of orbitals and magnetism.

\end{abstract}

\maketitle

Segregation at grain boundaries (GBs) affects various properties of polycrystals such as grain growth~\cite{Scott02,Cho97}, liquid metal embrittlement (LME)~\cite{Joseph99, Wolski08} and corrosion~\cite{Viswanadham80, Knight10}. However, the exact segregated structures, and hence the underlying mechanisms at atomic level, are far from being fully revealed. As a generalization of Gibbs' definition of phase, the new concept ``complexion'' was proposed to describe thermodynamically stable interfacial structures~\cite{Tang06, Harmer11}. A particular type of complexion, Dillon-Harmer complexions~\cite{Cantwell14}, attracted attention after the discovery of six discrete complexions and their striking connection with grain boundary kinetics in alumina~\cite{Dillon07}, a widely used ceramic material. 

Recently, Dillon-Harmer complexions were discovered in metallic systems Bi-Ni~\cite{Luo11} and Bi-Cu~\cite{Kundu13}, which could possibly explain the long standing puzzle of LME. In these experiments, Bi formed bilayer films ubiquitously in Ni at general orientation GBs around the penetration tip (as far as 983 $\mu m$). In contrast, a clean low energy Ni grain boundary was found 239 $\mu m$ away from the tip. Bi also formed bilayer films at Cu GBs around the penetration tip. However, the bilayer films were only observed close to the tip (as far as 257 $\mu m$) indicating bilayer films were stable over a much narrower Bi chemical potential window. Similarly to Ni, Bi did not segregate at low energy Cu GBs. A study of Fe revealed no Bi films~\cite{Luo13}. 

A recent theoretical study~\cite{Kang13} of Bi at Ni and Cu(111) twist and $\Sigma$5(310) GBs found the Bi bilayer enthalpy of formation on $\Sigma$5(310) is negative, which indicates thermodynamic stablity, while on (111) twist GBs it is positive. The authors proposed that bilayers are more stable than monolayers based on interaction strength between Bi and Ni layers and an electric dipole generated in the Bi bilayer on (111) twist GB. However, neither the origin of different segregation behavior of Bi on Ni compared with Cu, nor the relative stability of bilayer and trilayer films, was discussed. Moreover, a detailed study of the film structure, registry and bonding character is needed. 

In this letter, we present a first-principles study of Bi films on low energy $\Sigma$3(111) and high energy $\Sigma$5(210) transition metal GBs. Our results agree with the experimental result that bilayer films form on Ni and Cu high energy GBs but do not form on Fe GBs (see Supplemental Material~\cite{Gao13supp} for Fe). To compare the trends within TMs of similar structure, we analyzed Bi on FCC Co GBs. By exploiting the weak Bi interlayer interaction, we proposed a model that can be used to predict Bi bilayer stability on various Ni GBs with relatively simple surface calculations. Moreover, we analyze the difference between transition metals based on localization of orbitals and magnetization, and confirm this analysis with crystal orbital Hamilton populations (COHP)~\cite{Dronskowski93, Jepsen95} calculations. We also confirm the antibonding between Cu atoms induced by Bi segregation which was inferred in Reference~\cite{Duscher04}. Finally, we discuss the effects of Bi bilayers on grain growth and embrittlement.  

\begin{figure}[ht]
\includegraphics[scale=0.3,clip]{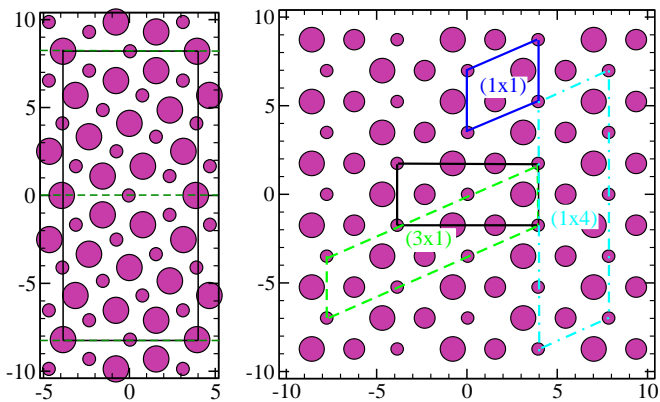}
\caption{(Color online) Left: Side view of our $\Sigma$5(012) GB. The black cell is our unit cell. The dashed green lines are GB planes. Right: Top view of $\Sigma$5(012) GB plane. Three layers of atoms are shown. The black solid cell is the orthorhombic unit cell we use to calculate $\Sigma$5(012) GB energies. Atom size indicates depth (large below small). Units are~\AA.}
\label{fig:Sig5-struc}
\end{figure}

Our calculation methods are similar to our study of Bi on Ni(111)~\cite{Gao13}, namely PAW potentials~\cite{Blochl94,Kresse99} in the PBE~\cite{Perdew96} generalized gradient approximation with default energy cutoffs using VASP~\cite{Kresse93,Kresse96}. To find stable structure at GBs, we first study Bi structures on free surfaces. For Bi on TM(111) and (120), we construct models based on four and six metal layers normal to the surface respectively with Bi films on one side. We choose the $\Sigma$3(111) twist and the $\Sigma$5(012) tilt GBs as representative low energy and high energy GBs respectively. $\Sigma$3(111) is formed by cleaving the bulk along the (111) plane, rotating one grain around [111] by 60$\rm ^0$ and rejoining the two parts~\cite{Sangid10}. It is the lowest energy GB among all coincidence site lattice (CSL) types that differ from bulk by just a stacking fault. $\Sigma$5(012) is formed by cleaving the bulk along the (012) plane, rotating one grain around [100] by 53.1$\rm ^0$ and rejoining the two parts after removing overlapping atoms (see Fig.~\ref{fig:Sig5-struc}).

For Bi on $\Sigma$3(111) GBs we stack six layers of metal with periodic boundary conditions and rotate three layers relative to the other three, thus creating the GBs. Then we insert our Bi film at one GB, leaving the other bare. To reduce computational complexity, the segregated structures at $\Sigma$5(012) GBs are calculated with six layers TM at each side of the Bi films and terminated by bare TM surfaces with vacuum at both sides. The $\Sigma$5(120) GB plane is shown in Fig.~\ref{fig:Sig5-struc}. Later on, we refer the blue solid cell as ($1 \times 1$), green dashed cell as ($3 \times 1$) and cyan dash-dotted cell as ($1 \times 4$). To analyze interaction strength and bonding character, we perform COHP calculations which evaluate matrix elements of the total energy between pairs of atomic orbitals on neighboring atoms. The differential (dCOHP) reveals the bonding and antibonding orbitals while the integral up to the Fermi energy (iCOHP) measures the bond strength.

Our calculated GB energies $E_{GB}$ are shown in Supplemental Material~\cite{Gao13supp}, Table S1, and agree well with prior literature. For Bi on the Ni(111) surface, we found a 4-atom Bi monolayer on a ($3 \times 3$) surface cell is stable over a wide Bi chemical potential~\cite{Gao13}, and the same holds true for Co. For Bi on the Cu(111) surface, 2-atom Bi monolayer on a [2012] cell is stable, which agrees with expermental observation~\cite{Kaminski05}. On TM(120) surfaces, Bi sitting on the valley sites of ($1 \times 1$) cells are stable over a wide range of chemical potential. This stable Bi on Cu(120) surface structure was observed in experiment~\cite{Blum01}. 

\begin{figure}[ht]
\includegraphics[scale=0.3,clip]{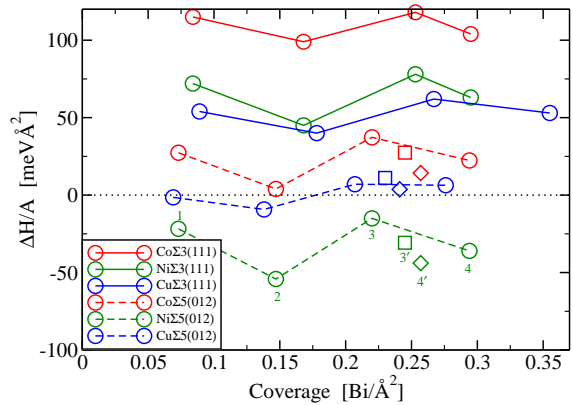}
\caption{(Color online) Enthalpies of Bi films at $\Sigma$3(111) and $\Sigma$5(012) GBs. Solid lines connect Bi films in ($3 \times 3$) cells of $\Sigma$3(111) GBs. Films of 1-3 layer thickness (labeled 1-3) have 4 Bi atoms per layer while the 4-layer films contain 4 Bi atoms per layer in layers adjacent to Ni but 3 Bi per layer in the middle two layers~\cite{Gao13} for Ni and Co. Films for Cu have 2 Bi per layer. Dashed lines connect Bi films in ($1 \times 1$) cells of $\Sigma$5(120) GBs. Red, green and blue colors indicate Co, Ni and Cu respectively. Square points (labeled as 3$'$) stand for trilayer films in ($3 \times 1$) cells of $\Sigma$5(012) GBs with denser middle layer (4 Bi in ($3 \times 1$) cell). Diamond points (labeled as 4$'$) stand for four layer films in ($1 \times 4$) cells of $\Sigma$5(012) GBs with the in-plane density of the middle bilayer similar to bulk Bi (3 Bi in a ($1 \times 4$) cell).}
\label{fig:enthalpy}
\end{figure}

We then study various Bi films at GBs. To compare the stability of these films, we calculate the enthalpy of formation, which is defined as,
\begin{equation} 
\label{eq:enthalpy}
\Delta H/A=[E_{\rm tot}-E^{\rm TM}_{\rm slab}-E^{\rm Bi}_{\rm bulk}N_{\rm Bi}]/A,
\end{equation} 
where $E_{\rm tot}$ is the energy of a TM slab containing GB segregated by Bi, $E^{\rm TM}_{\rm slab}$ is the energy of a TM slab containing a bare GB, $E^{\rm Bi}_{\rm bulk}$ is the Bi bulk energy, A is the GB area. Fig.~\ref{fig:enthalpy} shows our enthalpies of formation. On the low energy $\Sigma$3(111) GBs, the enthalpies of Bi film formation are all large and positive, which suggests that Bi does not form stable films at these GBs. This is expected since the $\Sigma$3(111) GB differ from bulk only by a low energy stacking fault. It is energetically unfavorable to cut the strong bulk-like metal bonds and replace them by bonds with Bi. These results agree with the experimental observation~\cite{Luo11, Kundu13} of bare Ni and Cu low energy GBs near the Bi penetration tip. At the high energy $\Sigma$5(120) GB, $\Delta$H is reduced for all TM. At Co $\Sigma$5(120) GB, $\Delta$H remains positive suggesting all films are unstable. For Ni, all Bi films have negative $\Delta$H which means Bi penetration is favorable for all these films. Moreover, bilayer Bi is most favorable, with lower enthalpy of formation than monolayer and trilayer. For Cu, Bi monolayer and bilayer film have negative enthalpies of formation. The bilayer preference is less pronounced than on Ni. Overall, the enthalpy of formation is less negative on Cu than on Ni, which indicates interfacial films are less favorable in the case of Cu. 

To further illustrate the stability of Bi films at $\Sigma$5(120) GBs, we calculate the GB free energy. This is the Legendre transformation of the enthalpy of formation (Eq.~\ref{eq:enthalpy}), replacing the Bi coverage with relative chemical potential. From equilibrium thermodynamics, the most stable structure at a certain Bi chemical potential minimizes the GB free energy $\gamma$~\cite{Kitchin08},
\begin{equation} 
\label{eq:gamma}
\gamma =[\Delta H-\Delta\mu_{\rm Bi}N_{\rm Bi}]/A,
\end{equation}
where $\Delta\mu_{Bi}\equiv\mu_{\rm Bi}-E^{\rm Bi}_{\rm bulk}$ is the Bi relative chemical potential. Note that $\Delta\mu_{\rm Bi}=0$ corresponds to the chemical potential of bulk Bi. 

\begin{figure}[ht]
\includegraphics[scale=0.3,clip]{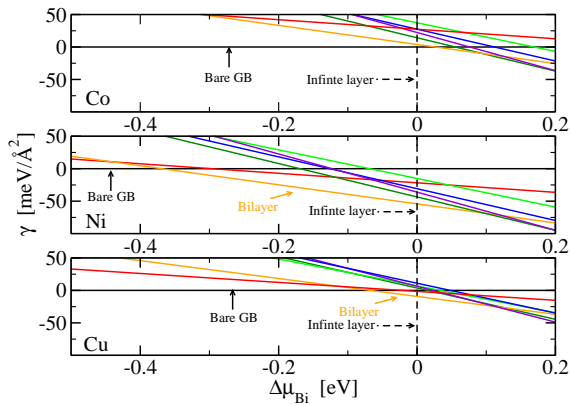}
\caption{(Color online) GB free energy of Bi films at Co, Ni, Cu $\Sigma$5(120) GBs, respectively, top to bottom. The black solid lines stand for bare GBs while the black dashed lines stand for infinite bulk-like Bi films. Other lines are for different Bi films, with stable bilayer labeled.}
\label{fig:line}
\end{figure}

As shown in Fig.~\ref{fig:line}, the stable sequence at Co $\Sigma$5(120) GB goes from a bare GB plane directly to an infinite height bulk-like film at $\Delta\mu_{Bi}=0$ eV. In contrast, the stable sequence at Ni $\Sigma$5(120) GB goes from bare GB to bilayer Bi film with ($1 \times 1$) cell at $\Delta\mu_{Bi}=-0.37$ eV, and then to the infinite height bulk like films at $\Delta\mu_{Bi}=0$ eV. The stable sequence at Cu $\Sigma$5(120) GB goes from bare GB to bilayer Bi film with ($1 \times 1$) cell at $\Delta\mu_{Bi}=-0.067$ eV, and then to the infinite height bulk-like films at $\Delta\mu_{Bi}=0$ eV. Bi films are thus not stable at Co $\Sigma$5(120) GB, and the Bi bilayer film at Ni $\Sigma$5(120) GB is stable over a much wider chemical potential window than at Cu $\Sigma$5(120) GB. These results are consistent with the experimental observations~\cite{Luo11, Kundu13} at high energy GBs of Ni and Cu.

\begin{figure}[ht]
\includegraphics[scale=0.5,clip]{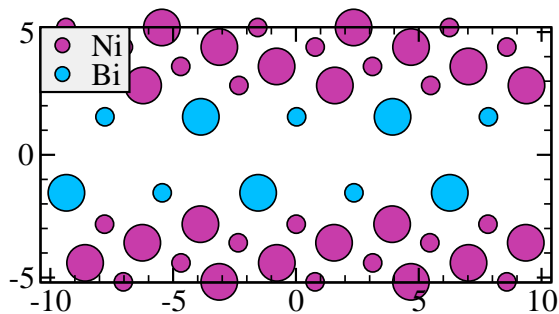}
\caption{(Color online) Relaxed monolayer and bilayer Bi films at ($1 \times 1$) cell of Ni $\Sigma$5(120) GB. Only Ni atoms close to Bi are shown. Atom size indicates depth (large below small). Units are~\AA.}
\label{fig:Bi-NiGB}
\end{figure}

Studying other bilayer films with different registry and coverage, it turns out the valley site of a missing Ni atom is a strong Bi adsorption site. For structures with Bi density smaller than the ($1 \times 1$) film, all Bi atoms relax into valley sites. The ($1 \times 1$) film is more stable than these films due to the energy gain by putting more Bi at the remaining empty valley sites. With Bi density larger than the ($1 \times 1$) film, Bi atoms in each layer bond with each other in the unfavorable metallic form and also leave some empty valley sites which weakens the bonds with Ni. Both these effects destabilize such films. This result is also consistent with our surface study that the ($1 \times 1$) monolayer structure is most stable. 

Trilayer films are unfavorable at all GBs, again because of the bonding character of Bi. Bulk Bi has the common $\alpha$-As group-V semimetal (strukturbericht A7, Pearson {\it hR2}) with rhombohedral space group $R\bar{3}m$ forming a bilayer structure. Each Bi atom has strong covalent bonds with three intrabilayer neighbors at the distance of 3.1~\AA~and bonds weakly with three interbilayer neighbors at 3.5~\AA. The trilayer films contain a chemically adsorbed monolayer on each side of the GB plus a monolayer of atoms in between that forms metallic bonds. The four layer structure has a bilayer film similar to the bulk structure between the strong adsorbed monolayer films. Thus, bilayer films and four layer films are more favorable than trilayer films. The observed trilayer Bi film at Ni GB near the penetration tip is thus indeed predicted to be a metastable structure as inferred in Reference~\cite{Luo11}. 

The Bi interlayer interaction is weak in the bilayer films at both Ni $\Sigma$3(111) and $\Sigma$5(120) GBs, with bond lengths around 3.9~\AA~and 4.2~\AA~respectively, which are larger than the weak Bi-Bi metallic bond. In experiment, the observed Bi layer spacing is 3.9$\pm$0.6~\AA. Based on these observations, we propose a model to calculate the enthalpy of formation of Bi bilayer at Ni GBs with bare GB energies and surface adsorptions, by neglecting the interlayer interaction,
\begin{equation}
\Delta H/A\approx E_{\rm surf}^a+\Delta H_{\rm ML}^a/A+E_{\rm surf}^b+\Delta H_{\rm ML}^b/A-E_{\rm GB} ,
\label{eqn:model}
\end{equation}
where $E_{\rm surf}^a$ and $E_{\rm surf}^b$ are the surface energies of Ni surfaces $a$ and $b$ adjacent to the GB plane, $ \Delta H_{\rm ML}^a$ and $\Delta H_{\rm ML}^b$ are the enthalpies of formation of Bi monolayers on Ni surfaces $a$ and $b$. The first four terms represent the excess energy per area with bilayer intercalation. $E_{\rm GB}$ is the excess energy per area without intercalation. We define $E _{\rm GB}^{\rm min}$ as the minimum energy of GB consisting of surfaces $a$ and $b$ such that formation of Bi bilayer is energetically favorable, i.e. for which $\Delta H/A\leq 0$ (see values in Tab. S2 in Supplemental Material~\cite{Gao13supp}). Hence
\begin{equation}
E_{\rm GB}^{\rm min}\approx E_{\rm surf}^a+\Delta H_{\rm ML}^a/A+E_{\rm surf}^b+\Delta H_{\rm ML}^b/A.
\end{equation}
Results of this model are given in Table S3 in Supplemental Material~\cite{Gao13supp}. The enthalpies of formation from model predictions ($\Delta H^{\rm model}$) and direct calculations ($\Delta H^{\rm calc}$) are within 0.01 eV/\AA$^2$, and slightly exceed the direct calculations because we neglect the interaction between Bi bilayers which lower the total energy. This model thus accurately predicts Bi bilayer enthalpies of formation, while being easier to calculate than direct Bi at Ni GBs. 

Approximate energies for many bare GBs can be obtained from empirical potential calculations~\cite{Olmsted09}. Based on those values, our model predicts that Bi bilayer enthalpies of formation are positive on all Ni(111) CSL twist GBs, but are negative on (100) CSL twist GBs for rotation angles between 10 and 50 degrees. Moreover, for structures where the optimized Bi monolayer is not commensurate with the GB cell, and for GBs with different adjacent surfaces (i.e. $a\neq b$) that are not commensurate with each other, the model prediction might be more accurate than affordable direct calculations. An example is the (111)/(100) GB in Table S3 in Supplemental Material~\cite{Gao13supp}, where a general GB is constructed with $a=$(111) and $b=$(100) planes with ($3 \times 3$) surface cell at the $a$ side and [2 -2 1 3] surface cell (following the notation of~\cite{Bollmann11}) at the $b$ side. Unlike the CSL GBs, the $\Delta H^{\rm calc}$ is greater than $\Delta H^{\rm model}$ due to strain of the Ni cells (around 5\%). This artificial strain introduced by forcing the two weakly interacting grains to share a common small cell makes the direct calculation inaccurate. This model could easily be generalized to other polycrystal materials providing the interlayer interaction of segregated films is small.

Ni GBs are severely embrittled by Bi bilayer segregation. In Table S4 in Supplemental Material~\cite{Gao13supp}, we show the work of separation (defined as $W_{\rm sep}= 2E_{\rm surf}-E_{\rm GB}$, the work needed to separate the GB~\cite{Finnis04}) at several bare and Bi segregated GBs. In all these GBs, $W_{\rm sep}$ is reduced by more than 95\% due to the weak interaction between Bi layers~\cite{Kang13}.

The differences of Bi interactions among these three transition metal GBs can be understood with a combination of localization of TM electron orbitals and magnetism. With increasing of atomic number from Co to Ni and to Cu, the 3d orbital becomes more localized, and thus the interactions between TMs and with Bi decreases. For example, shown in Tab. S4 in Supplemental Material~\cite{Gao13supp} the (012) surface energies $E_{\rm surf}$ and Bi monolayer $\Delta H/A$ on (012) surface diminish for Co, Ni and Cu respectively in nonmagnetic calculation. With magnetism, Co and Ni(012) $E_{\rm surf}$ decrease further due to increasing surface magnetic moment. The remaining greater Co surface energy due to stronger interaction between less localized orbitals makes the Co GBs harder to separate.

Values of iCOHP measure bond strength. The Bi-Co bonds at surface is weaker than Bi-Ni when magnetism is included (see Tab. S4 in Supplemental Material~\cite{Gao13supp}) due to the fact that Bi is nonmagnetic and quenches the TM surface magnetic moments. All of these effects are greater at the Co surface than Ni due to Co's larger surface magnetic moments, 1.93$\mu_B$/atom compared with 0.78$\mu_B$/atom for Ni. This leads to a greater increase in Bi $\Delta H/A$ on Co(012) than Ni(012) surface, compared with the nonmagnetic case. This effect is also manifested from our COHP calculation that the Bi-Co bond is weakened by 0.12 eV while Bi-Ni bond is weakened only by 0.02 eV due to magnetism. Thus Bi monolayers on Co(012) surface and Bi bilayers on Co $\Sigma$5(012) GB are less favorable to form than on Ni due to the stronger interaction between Co atoms than between Ni atoms resulting from greater localization of 3d electrons on Ni, and weaker interaction between Bi and Co than between Bi and Ni, due to magnetism. 

Apart from weaker interaction of Bi with Cu than with Ni, the Bi bilayer is less favorable on Cu than on Ni, since Bi gives electrons to Cu, half filling the Cu $s$ states. This leads to stronger antibonding among Cu atoms close to Bi as is inferred in~\cite{Duscher04} and confirmed by our COHP calculation shown in Fig. S2 in Supplemental Material~\cite{Gao13supp}. The Ni-Ni and Co-Co bonds close to Bi however are stronger than in the bulk, where no antibonds appear. 

In conclusion, we have studied Bi segregation at Co, Ni and Cu low energy $\Sigma$3(111) and high energy $\Sigma$5(120) GBs using density functional theory. Our results reproduce the experimental result that Bi does not form film at all Fe GBs but forms a bilayer film ubiquitously at Ni high energy GBs, and in a much narrower chemical potential window at Cu high energy GB. The difference between these metals can be explained by the localization of 3d orbitals and also the loss of magnetism near the GB of Co (and presumably Fe). Moreover, Bi on Cu GB also increases the strength of antibonding, as confirmed by COHP calculation. We propose a model to predict the stability of Bi bilayer at various Ni GBs. Combining with the empirical GB energies from Reference~\cite{Olmsted09}, the model suggests Bi bilayer is not thermodynamically stable on (111) twist CSL GBs but should be stable in most (100) twist CSL GBs. 

LME is a long standing puzzle that has not been well understood at an atomic level. Several studies based on monolayer adsorption propose explanations of atomic size effects~\cite{Finnis04} or electronic effects~\cite{Duscher04}. Our theoretical study confirms the formation of Bi bilayers as the origin of the LME of Ni. The Bi bilayer (instead of monolayer) film at Ni and Cu GB thus serves as a starting point for future study of LME. 

Segregation also affects grain growth by altering the GB energy and mobility. When GB energy drops, grain growth is reduced since the driving force for grain growth is proportional to GB energy. Indeed, some metastable structures with low but positive GB energy halt grain growth. W dopants at Ni GB stops the grain growth when the averaged GB energy drops by $\sim$60\%~\cite{Detor07}. We find Bi bilayer segregation at Ni $\Sigma$5(120) GB reduces the GB energy from 77 meV/\AA$^2$ to 28 meV/\AA$^2$, a reduction of 64\%. We thus anticipate a significant change in the grain growth behavior for Bi segregated Ni polycrystal since Bi bilayer films are ubiquitous at general GBs.

The authors thank Jian Luo, Jeffrey Rickman, Zhiyang Yu, Anthony Rollett, Gregory Rohrer and Martin Harmer for helpful discussion. Financial support from the ONR-MURI under the grant NO. N00014-11-1-0678 is gratefully acknowledged.

\bibliographystyle{apsrev}
\bibliography{BiNi}

\end{document}


\beginsupplement
\title{Supplemental Material for:\\ First Principles Study of Bismuth Films at Transition Metal Grain Boundaries}
\author{Qin Gao and Michael Widom}
\affiliation{Department of Physics, Carnegie Mellon University, Pittsburgh, Pennsylvania 15213, USA}

\maketitle

\section{Bi bilayer on Fe $\Sigma$5(012) GB}
\begin{figure}[ht]
\includegraphics[scale=0.6,clip]{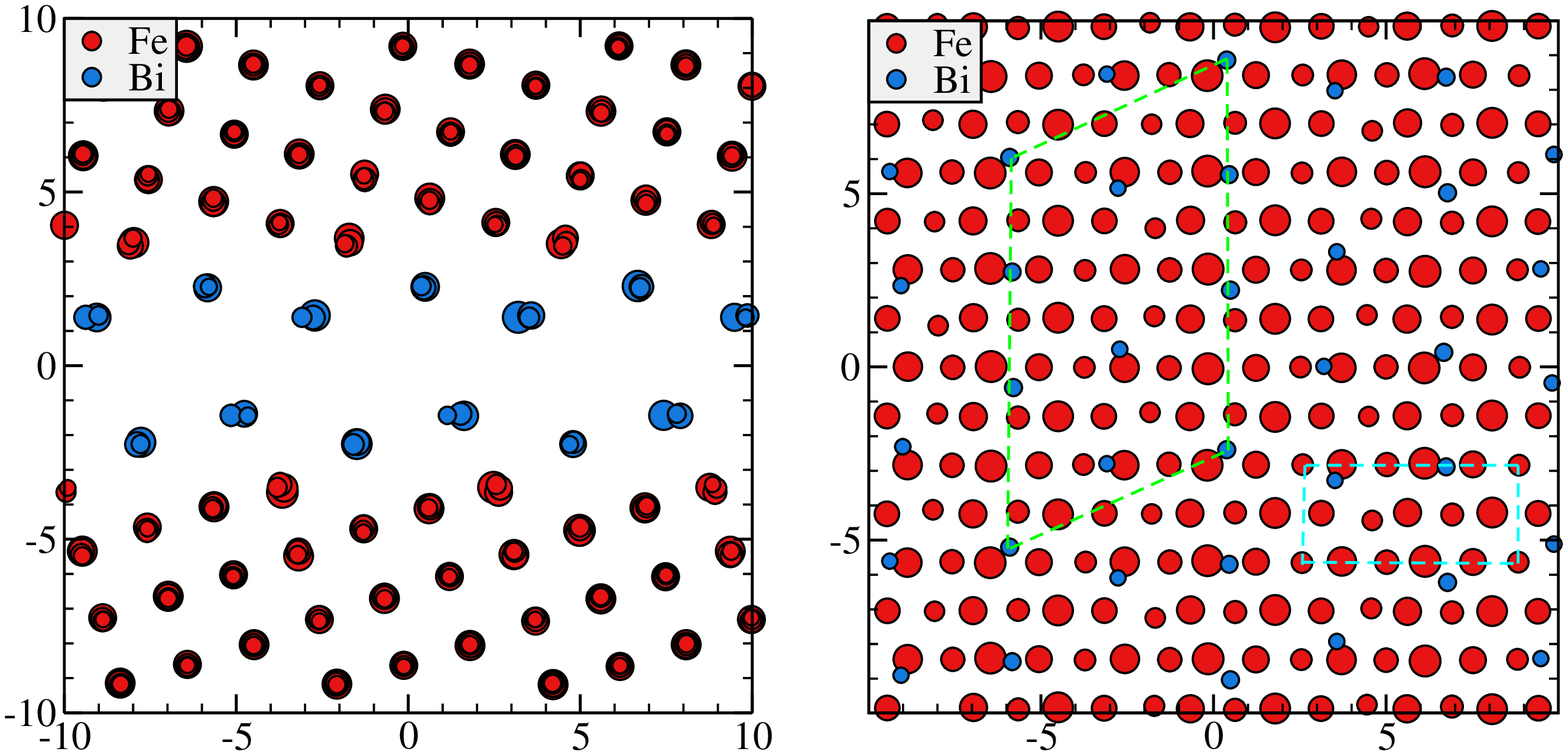}
\caption{(Color online) Side view (left) of relaxed Bi bilayer at Fe $\Sigma$5(012) GB and top view (right) of Bi monolayer on one side of the GB plane. The cyan cell is the Fe GB unit cell, the green cell is the Bi segregated GB unit cell. Atom size indicates depth (large below small). Length units are in~\AA.}
\label{fig:supp-Fe}
\end{figure}

We calculated Bi bilayer enthalpy on Fe $\Sigma$5(012), a high energy GB which is created by cleaving the BCC bulk along the (012) plane and rotating one grain around [001] by 53.1$^o$ and rejoining the two parts. Our calculated GB energy is 98 meV/\AA$^2$, close to the GB energy 104 meV/\AA$^2$ of the lowest energy structure in the literature~\cite{Ossowski10}. We first studied Bi monolayers on Fe (012) surfaces and then calculated bilayer films on the GB with the stable surface structure at two sides of the GB plane. We used 10 layers of Fe at each side of the Bi film. The relaxed structure is shown in Fig.~\ref{fig:supp-Fe}. The lowest $\Delta$H/A is +17 meV/\AA$^2$ for bilayer films, which is large and positive indicating that the Bi bilayer is not stable even on this high energy GB.

\begin{table}[ht]
\begin{tabular}{|l|r|r|r|}
\hline
    ~~~GB         & ~~~~~~~Co~~~~~~~~~& ~~~~~~Ni~~~~~~~~~& ~~~~~~~~Cu~~~~~~~~ \\
\hline
$\Sigma$ 3(111)&~~-1.6~~~~~~~~ & ~2.8 (2.7~\cite{Murr07})~&0.9 (1.4~\cite{Murr07})~~~\\
\hline
$\Sigma$ 5(120)&~80~~~~~~~~ &~77 (89~\cite{Yamaguchi04})~~&55 (59~\cite{Tschopp07})~~~ \\
\hline
\end{tabular}
\caption{\label{tab:GBenergy} GB energies, units are meV/\AA$^2$. The energy conversion factor is 1 meV/\AA$^2=$0.016 J/m$^2$. Values from other studies are in parentheses. Note the Co $\Sigma$3(111) GB energy is negative because the $T=0$ K state is HCP rather than the high-T FCC that we choose to compare with.}
\end{table}

\begin{table}[ht]
\begin{tabular}{|l|r|r|r|}
\hline
~Surface~~&~$E_{\rm surf}$~~&~$\Delta H_{\rm ML}$/A~~&~$E_{\rm GB}^{\rm min}$ ~\\
\hline
~~(111)&~0.118~~&-0.090~~~~~&0.056~~\\
\hline
~~(001)&~0.137~~&-0.122~~~~~&0.030~~\\
\hline
~~(120)&~0.150~~&-0.133~~~~~&0.034~~\\
\hline
\end{tabular}
\caption{\label{tab:surf} Calculated input quantities for the enthalpy model (Eq. 3). Predicted $E_{\rm GB}^{\rm min}$ values for GB with the same surface plane (i.e. $a=b$) at two sides. The energy units are eV/\AA$^2$. The Bi monolayer structure on Ni(100) surface is the $c$($2 \times 2$) structure as observed in experiment~\cite{Panja00}.}
\end{table}

\begin{table}[ht]
\begin{tabular}{|l||r|r|r||r|r|r|}
\hline
~~~~~GB& $E_{\rm GB}$ ~&~$\Delta H^{\rm model}$/A ~& ~$\Delta H^{\rm calc}$/A ~& ~$W_{\rm sep}^{\rm bare}$~ & ~$W_{\rm sep}$~~& Reduction\\
\hline
~~$\Sigma$3(111)~&~0.003~~&~0.053~~~~~~&~0.045~~~~~& 0.235~~~& 0.009~~& 96.2\%~~~~\\
\hline
~~$\Sigma$7(111)~&~0.029~~&~0.027~~~~~~&~0.023~~~~~& 0.209~~~& 0.009~~& 95.7\%~~~~\\
\hline
~~$\Sigma$5(100)~ &~0.064~~&~-0.034~~~~~~&~-0.037~~~~~& 0.208~~~& 0.007~~& 96.6\%~~~~\\
\hline
~~$\Sigma$5(120)~ &~0.077~~&~-0.043~~~~~~&~-0.054~~~~~& 0.220~~~& 0.010~~& 95.5\%~~~~\\
\hline
(111)/(100)&~0.055~~&~-0.012~~~~~  &~-0.004~~~~~& 0.207~~~& 0.004~~& 98.0\%~~~~\\
\hline
\end{tabular}
\caption{\label{tab:GB} Model and calculated Ni GB energies and Bi bilayer enthalpies of formation at different Ni GBs. Work of separation for bare and Bi bilayer segregated GBs is shown on right. The energy units are eV/\AA$^2$. The $\Sigma$7(111) GB is made by twisting the one side of bulk Ni by 21.8$\rm ^o$ around the [111] axis with (111) as GB plane. The resulting GB cell is [3-112] as defined in~\cite{Bollmann11} on which Bi favors 3 atoms per layer. The $\Sigma$5(100) GB is made by twisting one side of bulk Ni by 36.9$^o$ around the [001] axis with (001) as GB plane.}
\end{table}

\begin{table}[ht]
\begin{tabular}{|l|r|r|r|r|r|}
\hline
   & Co(nonmag) ~&~Co(mag)~ & ~Ni(nonmag)~ & ~Ni(mag)~ & ~Cu~~~~\\
\hline
~E$_{\rm surf}$ (eV/\AA$^2$)~&~0.193~~~~~~&~0.164~~~~&~0.152~~~~~~& 0.150~~~~& ~0.100~~~\\
\hline
~$\Delta$H (eV/\AA$^2$)~&~-0.160~~~~~~&~-0.118~~~~&~-0.152~~~~~~& -0.133~~~~& ~-0.070~~~\\
\hline
~iCOHP(Bi-TM)~&~-1.75 ~~~~~~&~-1.63~~~~~&~ -1.77~~~~~~~& -1.75~~~~~& ~-1.33~~~~\\
\hline
~iCOHP(TM-TM)$^a$~&~-1.38~~~~~~~&~-1.32~~~~~&~~~-1.13 ~~~~~~& -1.13~~~~~& ~-0.43~~~~\\
\hline
~iCOHP(TM-TM)$^b$&~-1.17~~~~~~~&~-1.16~~~~~&~-0.85~~~~~~~& -0.83~~~~~& ~-0.66~~~~\\
\hline
\end{tabular}
\caption{\label{tab:bonds} The (012) surface energy, Bi monolayer enthalpies of formation, integrated COHP (iCOHP) energies of Bi-TM bond, TM-TM bond near to Bi(a) and TM-TM bond away from impurities(b). The energy units are eV/bond for the iCOHP energies.}
\end{table}

\begin{figure}[ht]
\includegraphics[scale=0.65,clip]{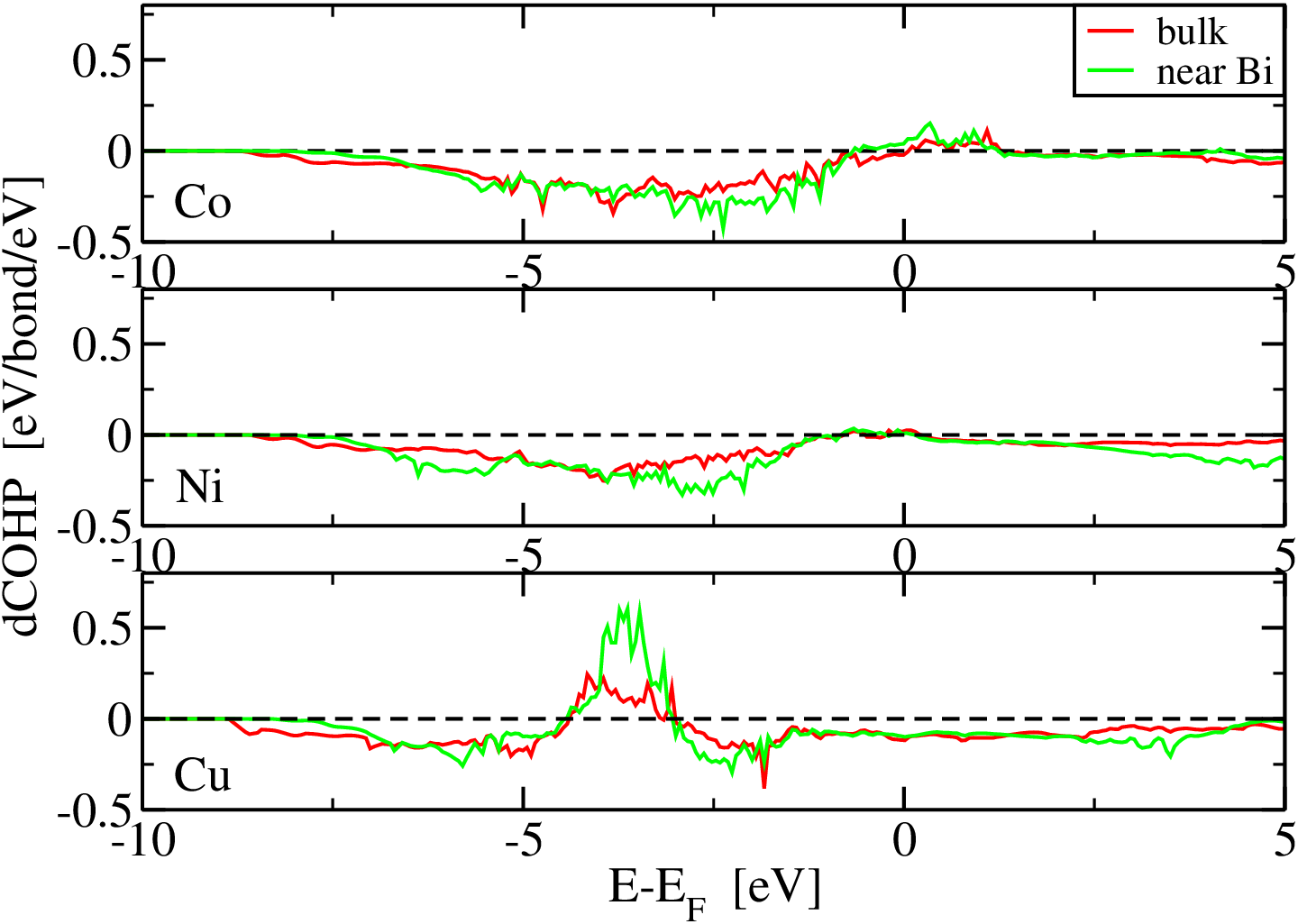}
\caption{(Color online) Differential COHP of Metal-Metal interaction in the bulk and near to Bi. Negative is bonding while positive is antibonding. The Ni and Co results are the summation of two spin components. The dashed green line is the x axis. The zero in x axis is the Fermi energy.}
\label{fig:dCOHP}
\end{figure}

\FloatBarrier
\bibliographystyle{apsrev}
\bibliography{BiNi}